\documentclass{moriond}

\usepackage[utf8x]{inputenc}
\usepackage{amsmath}
\usepackage{amssymb}

\newcommand{\Photo}{}

\begin{document}
\vspace*{4cm}
\title{Constraining the MSSM Higgs sector using precise Higgs mass predictions}

\author{ Henning Bahl }

\address{DESY, Notkestraße 85,\\
D-22607 Hamburg, Germany}

\maketitle\abstracts{
Different approaches are used for the calculation of the SM-like Higgs boson mass in the MSSM: the fixed-order diagrammatic approach is accurate for low SUSY scales; the EFT approach, for high SUSY scales. Hybrid approaches, combining fixed-order and EFT calculations, allow to obtain a precise prediction also for intermediary SUSY scales. Here, we briefly discuss the hybrid approach implemented into the code \texttt{FeynHiggs}. In addition, we show how the refined Higgs mass prediction was used to define new MSSM Higgs benchmark scenarios.}

%%%%%%%%%%%%%%%%%%%%%%%%%%%%%%%%%%%%%%%%%%%%%%%%%%

\section{Introduction}

It is a special feature of the MSSM that it allows to predict the mass of the SM-like Higgs boson, $M_h$, in terms of the model parameters. At the tree-level, it is determined by only two non-SM parameters, conventionally chosen to be the mass of the $\mathcal{CP}$-odd Higgs boson $A$, $M_A$, and the ratio of the vacuum expectation values of the two Higgs doublets, $\tan\beta$. Loop corrections, depending also on the parameters of other sectors, lead to a significant upwards shift of the tree-level value.

Currently, there a three different approaches for the calculation of the quantum corrections. In the most straightforward fixed-order approach, corrections to the Higgs self-energies are evaluated by calculating Feynman diagrams at a specific order in the full MSSM. This approach has the advantage that all terms at a specific order are incorporated. In case of a large hierarchy between the electroweak and the SUSY scale, however, large logarithms appear in the calculation exacerbating the convergence of the perturbative series. These logarithms can be resummed in the effective field theory (EFT) approach. In its simplest form all SUSY particles are integrated out at a single scale such that below this SUSY scale the SM is recovered as EFT. Matching conditions at the SUSY scale fix the SM Higgs self-coupling. RGE running down to the electroweak scale, where the physical Higgs mass is calculated, corresponds to a resummation of large logarithms. Due to this resummation, the EFT approach is precise for high SUSY scales. Typically, no higher dimensional operators are included in the EFT below the SUSY scale. Therefore, terms suppressed by the SUSY scale are neglected. Thus, the EFT calculation looses its accuracy in case of a low SUSY scale.

To obtain a precise precise prediction also for intermediary SUSY scales, the diagrammatic and the EFT approach can be combined~\cite{Hahn:2013ria,Bahl:2016brp,Athron:2016fuq,Staub:2017jnp,Bahl:2017aev,Athron:2017fvs,Bahl:2018jom,Bahl:2018ykj}. Here, we will discuss the hybrid approach implemented into the publicly available code \texttt{FeynHiggs}~\cite{Heinemeyer:1998yj,Heinemeyer:1998np,Degrassi:2002fi,Frank:2006yh,Hahn:2009zz,Hahn:2013ria,Bahl:2016brp,Bahl:2017aev,Bahl:2018qog}. Afterwards, we show how this calculation has been used to define new MSSM Higgs benchmark scenarios~\cite{Bahl:2018zmf,Bahl:2019ago}.

%%%%%%%%%%%%%%%%%%%%%%%%%%%%%%%%%%%%%%%%%%%%%%%%%%
%%%%%%%%%%%%%%%%%%%%%%%%%%%%%%%%%%%%%%%%%%%%%%%%%%

\section{Calculating the SM-like Higgs boson mass}
\label{sec:MhCalc}

The result of the fixed-order (FO) approach are the renormalized Higgs self-energies in the full MSSM. We denote the self-energy of the SM-like Higgs boson by $\hat\Sigma_{hh}^{\rm{FO}}(p^2)$, where $p^2$ is the external momentum. The result of the EFT approach is encoded in the SM Higgs tree-level mass, $2\lambda(M_t)v^2$, which is obtained in terms of the SM Higgs self-coupling $\lambda$ evaluated at the electroweak scale multiplied with the SM Higgs vacuum expectation value, $v$.

The basic idea of the hybrid approach implemented in \texttt{FeynHiggs} is to add the non-logarithmic terms of the fixed-order result (including all terms suppressed by the SUSY scale) and the resummed logarithms obtained in the EFT approach,
\begin{eqnarray}
\hat\Sigma_{hh}^{\rm{hybrid}}(p^2) = \hat\Sigma_{hh}^{\rm{FO}}(p^2)\Big|_{\rm{non-log}} -  2\lambda(M_t)v^2\Big|_{\rm{log}}.
\end{eqnarray}
We obtain the non-logarithmic terms of the Higgs self-energy by subtracting all contained logarithms; and the logarithmic terms of the EFT result, by subtracting all contained non-logarithmic terms. This improved self-energy is then used to obtain the physical Higgs mass by solving the Higgs pole equation within the full MSSM.

%%%%%%%%%%%%%%%%%%%% figure %%%%%%%%%%%%%%%%%%%%%%
\begin{figure}
\begin{minipage}{0.46\textwidth}
\includegraphics[width=\textwidth]{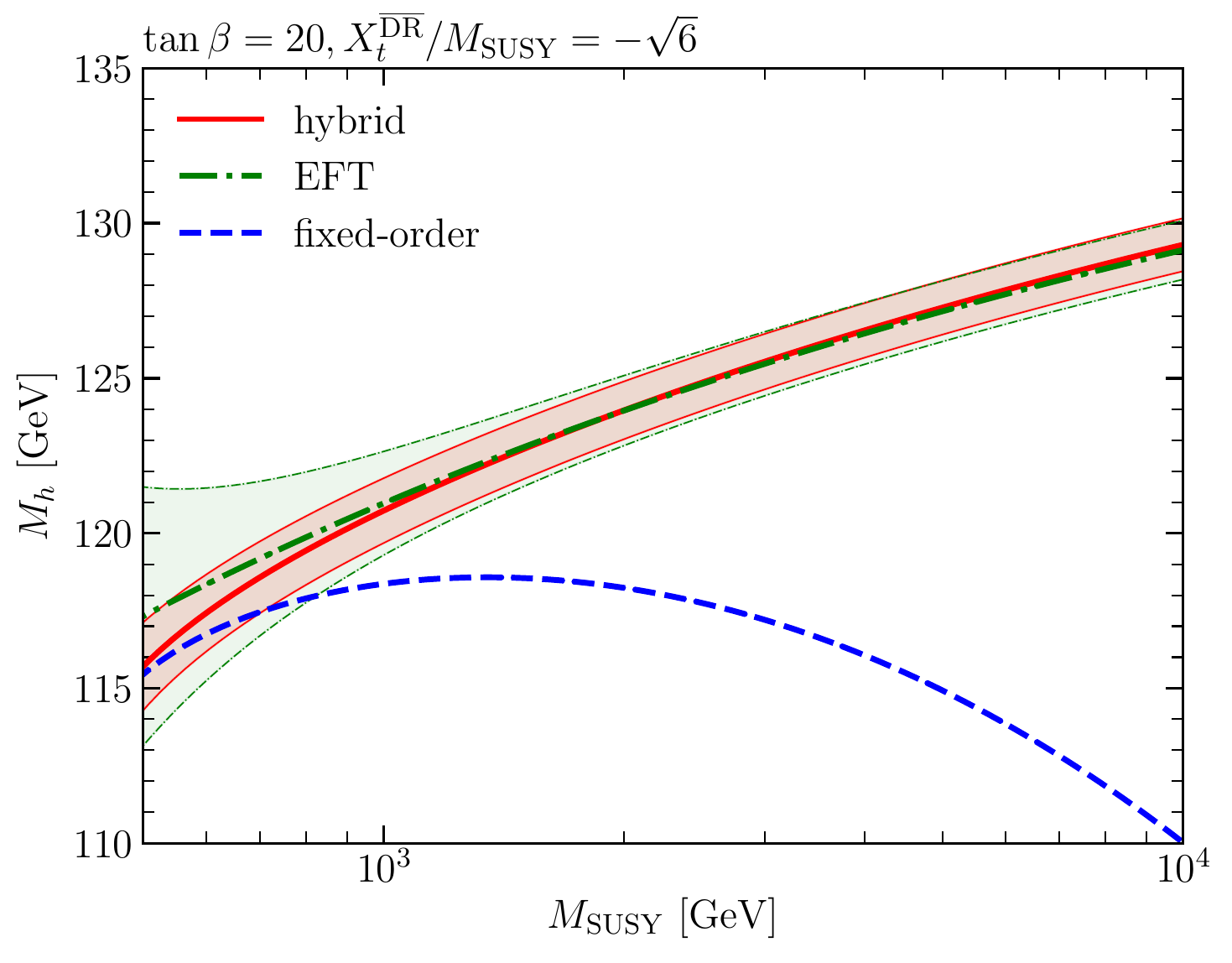}
\end{minipage}
\begin{minipage}{0.54\textwidth}
\includegraphics[width=\textwidth]{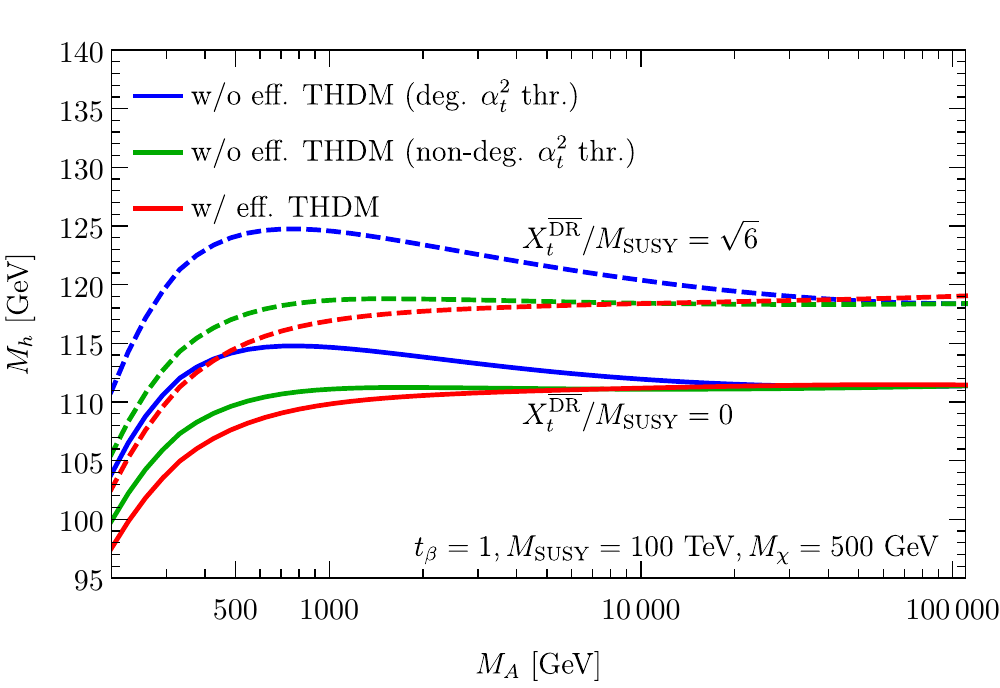}
\end{minipage}
\caption[]{Left: Results of the hybrid approach (red solid), the EFT approach (green dot-dashed) and the fixed-order approach (blue dashed) for the mass of the SM-like Higgs boson, $M_h$, in dependence of the SUSY scale, $M_{\rm{SUSY}}$. The colored bands depict the associated estimate of the remaining theoretical uncertainties. Right: Results of the hybrid approach for $M_h$ in dependence of $M_A$. As EFT below the SUSY scale, we use either the THDM (red) or the SM, employing degenerate (blue) or non-degenerate (green) threshold corrections. The results are shown for $X_t^{\overline{\rm{DR}}}/M_{\rm{SUSY}}=0$ (solid) and $X_t^{\overline{\rm{DR}}}/M_{\rm{SUSY}}=\sqrt{6}$ (solid).}
\label{fig:MhCalc}
\end{figure}
%%%%%%%%%%%%%%%%%%%% figure %%%%%%%%%%%%%%%%%%%%%%

In this way, the final result for the physical result for the SM-like Higgs boson mass includes all logarithms resummed in the EFT approach as well as all suppressed terms contained in the fixed-order approach. Therefore, the result of the hybrid approach should approach the fixed-order result for low SUSY scales and the EFT result for high SUSY scales. This is shown in the left plot of Fig.~\ref{fig:MhCalc} depicting a simplified scenario with the masses of all non-SM particles fixed to the SUSY scale setting the stop mixing parameter $X_t^{\overline{\rm{DR}}}=-\sqrt{6}M_{\rm{SUSY}}$ and $\tan\beta=20$. For low SUSY scales, where terms suppressed by the SUSY scale are relevant, the hybrid approach is in good agreement with the fixed-order result, whereas the EFT result, which does not include suppressed terms, yields a larger result for $M_h$. For large SUSY scales, where the EFT approach is precise, the hybrid approach and the EFT approach are in good agreement, whereas the fixed-order calculation yields a much lower result. The theoretical uncertainty estimate for the hybrid result~\cite{Bahl:2019xxx} is smaller or equal in size as the estimate for the EFT calculation.

So far, we have only considered the simple case of the SM as EFT below the SUSY scale. This choice is not appropriate if some of the non-SM particles are much lighter than the SUSY scale. We study the case of light non-SM like Higgs bosons in the right plot of Fig.~\ref{fig:MhCalc} choosing ${M_{\rm{SUSY}}=100}$ TeV, $\tan\beta=1$ (the electroweakino mass scale, $M_\chi$, is set to 500 GeV). For both considered values of the stop mixing parameter, the use of the THDM, with added electroweakinos,~\cite{Bahl:2018jom} amounts to a downwards shift of $\sim 2$ GeV in comparison to the calculation employing the SM as EFT if $M_A\ll M_{\rm{SUSY}}$. If in the SM case, $M_A$ is set equal to $M_{\rm{SUSY}}$ in the two-loop threshold correction for $\lambda$ controlled by the top Yukawa coupling, this shift is enlarged to up to 8 GeV. If $M_A \sim M_{\rm{SUSY}}$, the different calculations are as expected in good agreement.

%%%%%%%%%%%%%%%%%%%%%%%%%%%%%%%%%%%%%%%%%%%%%%%%%%
%%%%%%%%%%%%%%%%%%%%%%%%%%%%%%%%%%%%%%%%%%%%%%%%%%

\section{Higgs benchmark scenarios}

The large number of free parameters in the MSSM prevents an easy interpretation of the measured properties of the SM-like Higgs boson as well as searches for additional Higgs bosons. Therefore, Higgs benchmark scenarios have been developed. In these scenarios, only two parameters are varied, typically chosen to be $M_A$ and $\tan\beta$. The other parameters are fixed such that one of the Higgs bosons is SM-like with each scenario featuring a distinct Higgs phenomenology. The progress achieved in the calculation of the SM-like Higgs boson mass, including the methods presented in Section~\ref{sec:MhCalc}, ruled out almost the whole parameter space of the original benchmark scenarios~\cite{Carena:1999xa,Carena:2002qg,Carena:2000ks,Carena:2013ytb,Carena:2015uoe}. Therefore, new benchmark scenarios were proposed~\cite{Bahl:2018zmf,Bahl:2019ago} using the most recent version of \texttt{FeynHiggs} (Higgs masses and branching ratios), \texttt{SusHi}~\cite{Harlander:2012pb,Harlander:2016hcx} (Higgs production cross-sections), \texttt{HiggsBounds}~\cite{Bechtle:2008jh,Bechtle:2011sb,Bechtle:2013wla,Bechtle:2015pma} (searches for additional Higgs boson) and \texttt{HiggsSignals}~\cite{Bechtle:2013xfa} (properties of the SM-like Higgs boson).

%%%%%%%%%%%%%%%%%%%% figure %%%%%%%%%%%%%%%%%%%%%%
\begin{figure}
\begin{minipage}{0.5\textwidth}
\includegraphics[width=\textwidth]{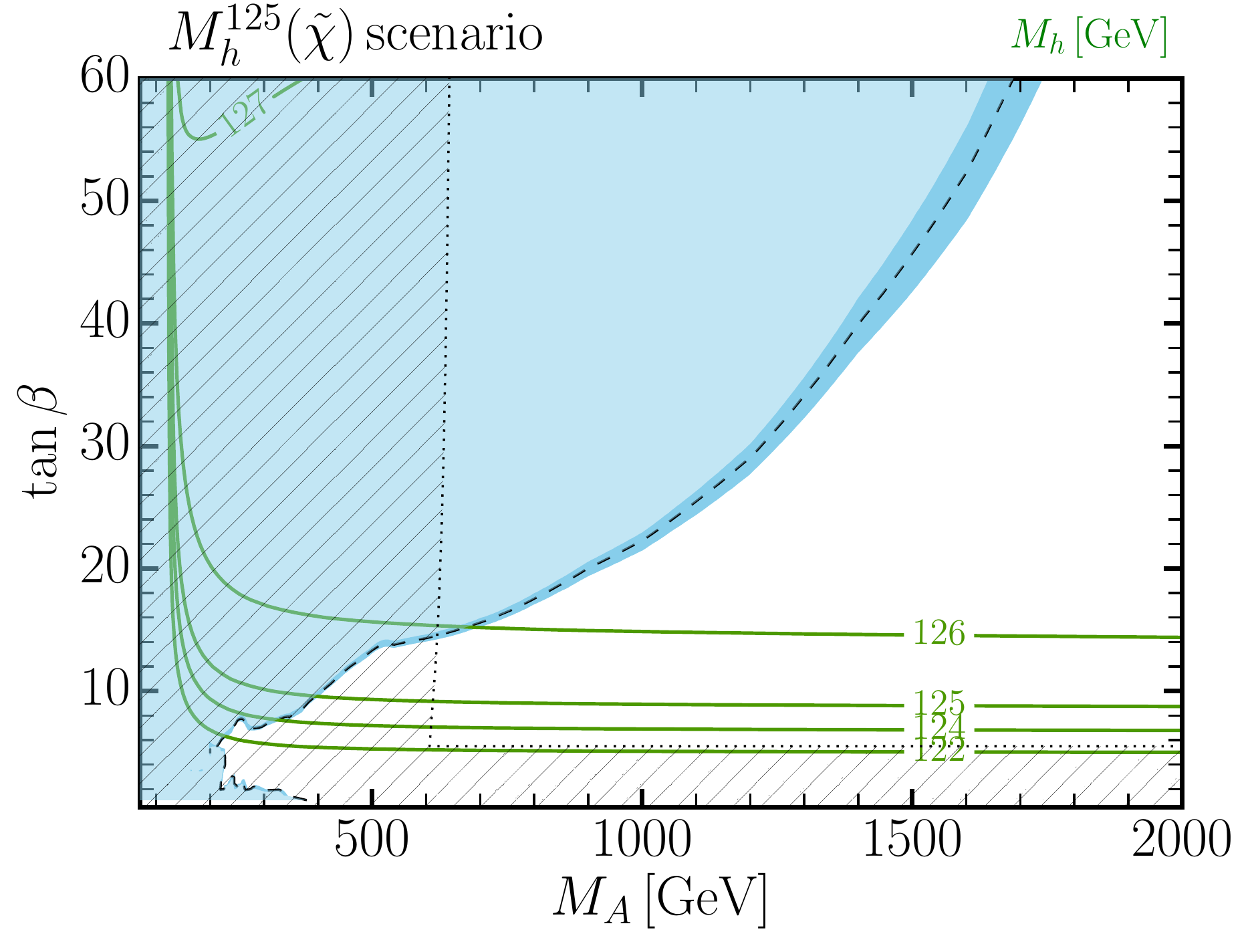}
\end{minipage}
\begin{minipage}{0.5\textwidth}
\includegraphics[width=\textwidth]{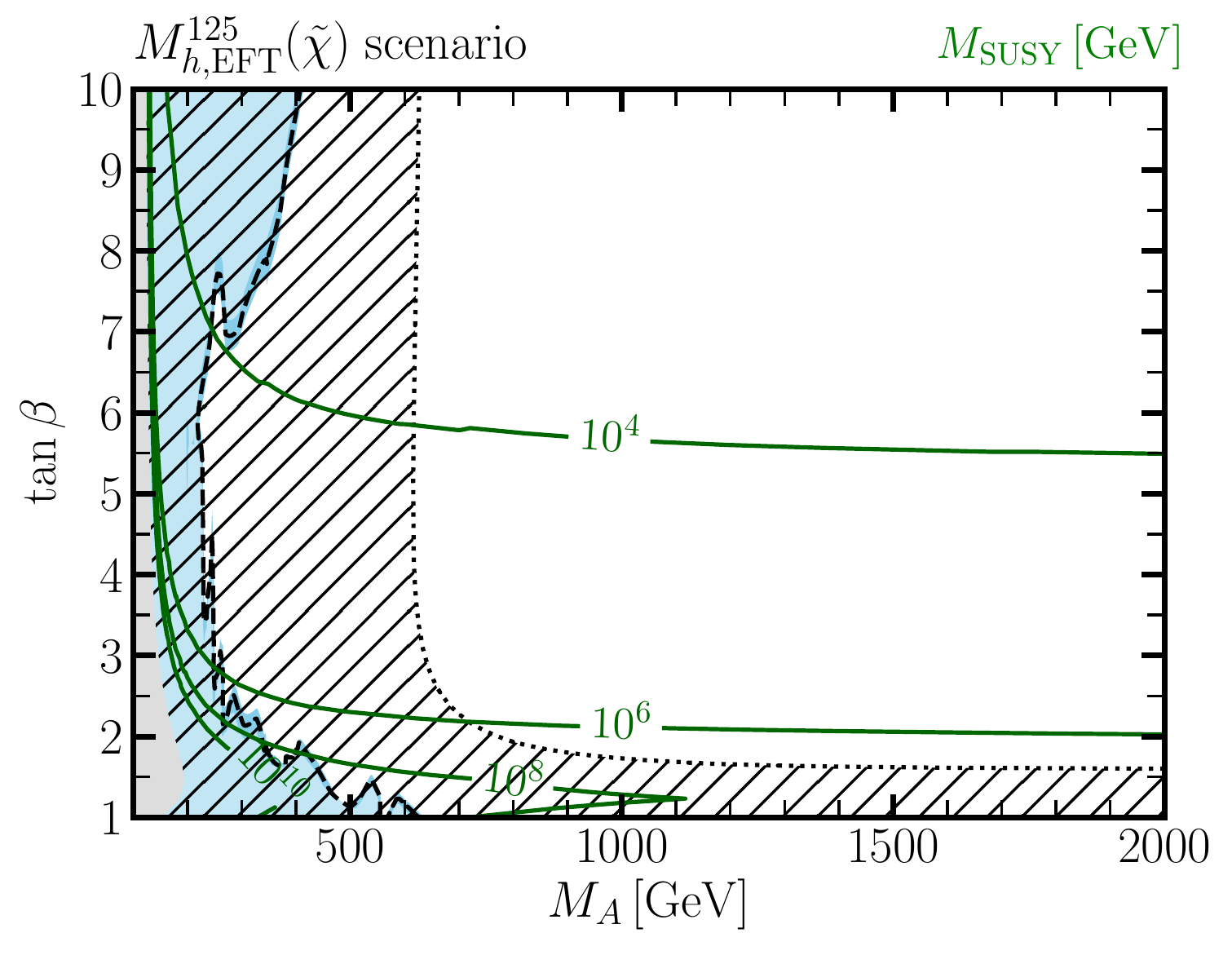}
\end{minipage}
\caption[]{Left: The $M_h^{125}(\tilde\chi)$ scenario shown in the $(M_A,\tan\beta)$ plane. The blue region indicates the area excluded by searches for additional Higgs bosons; the hashed region, the area excluded by measuring the properties of the SM-like Higgs boson. The green contours indicate the mass of the SM-like Higgs boson. Right: Same as left plot for the $M_{h,\rm{EFT}}^{125}(\tilde\chi)$ scenario, but the green contours indicate the required SUSY scale. In the grey area, $M_h \sim 125$~GeV can not be reached without raising $M_{\rm{SUSY}}$ above the upper limit of $10^{16}$ GeV.}
\label{fig:benchmarkscenarios}
\end{figure}
%%%%%%%%%%%%%%%%%%%% figure %%%%%%%%%%%%%%%%%%%%%%

In Fig.~\ref{fig:benchmarkscenarios}, we show the $M_h^{125}(\tilde\chi)$ (left plot) and the $M_{h,\rm{EFT}}^{125}(\tilde\chi)$ (right plot) scenarios as examples. Both scenarios feature light electroweakinos with masses $\sim 200$ GeV. The masses of the other SUSY particles are chosen above the TeV scale. In the $M_h^{125}(\tilde\chi)$ scenario the stop mass scale is fixed to 1.5 TeV. Therefore, the mass of the SM-like Higgs boson is too low for $\tan\beta\lesssim 6$. This parameter space is reopened in the $M_{h,\rm{EFT}}^{125}(\tilde\chi)$ scenario by adjusting $M_{\rm{SUSY}}$ at every point in the plane such that $M_h \sim 125$ GeV (with an upper limit of $10^{16}$ GeV).

In both scenarios, the measurements of the Higgs signal strengths lead to a lower limit for $M_A$ of $\sim 600$ GeV. In the $M_h^{125}(\tilde\chi)$ scenario the lower limit on $M_A$ increase with raising $\tan\beta$ due to constrains from direct searches for neutral heavy Higgs bosons decaying to a pair of tau leptons. This decay mode is irrelevant in the $M_{h,\rm{EFT}}^{125}(\tilde\chi)$ scenario. In this scenario, however, the presence of light charginos leads to an enhancement of the SM-like Higgs to $\gamma\gamma$ decay for low $\tan\beta$ allowing to exclude the region of $\tan\beta \lesssim 1.5$.

%%%%%%%%%%%%%%%%%%%%%%%%%%%%%%%%%%%%%%%%%%%%%%%%%%
%%%%%%%%%%%%%%%%%%%%%%%%%%%%%%%%%%%%%%%%%%%%%%%%%%

\section{Conclusion}

We discussed how the fixed-order and EFT approaches for the calculation of the SM-like MSSM Higgs boson mass can be combined. The resulting hybrid approach allows a precise prediction of the Higgs boson mass for low, intermediary and high SUSY scales. In addition, we showed how the improved calculation was used as an important constrain for the definition of new MSSM Higgs benchmark scenarios.

%%%%%%%%%%%%%%%%%%%%%%%%%%%%%%%%%%%%%%%%%%%%%%%%%%
%%%%%%%%%%%%%%%%%%%%%%%%%%%%%%%%%%%%%%%%%%%%%%%%%%

% \section*{Acknowledgments}
%
% I would like to thank my collaborators which were involved in the projects discussed above.

%%%%%%%%%%%%%%%%%%%%%%%%%%%%%%%%%%%%%%%%%%%%%%%%%%
%%%%%%%%%%%%%%%%%%%%%%%%%%%%%%%%%%%%%%%%%%%%%%%%%%

% \section*{Appendix}

%%%%%%%%%%%%%%%%%%%%%%%%%%%%%%%%%%%%%%%%%%%%%%%%%%
%%%%%%%%%%%%%%%%%%%%%%%%%%%%%%%%%%%%%%%%%%%%%%%%%%

\end{document}